\documentclass{elsarticle}
\usepackage{latexsym}
\usepackage{float}
\usepackage[latin2]{inputenc}
\usepackage{graphicx}
\usepackage{ae}
\usepackage{aecompl}
\usepackage{setspace}
\usepackage{amsmath}
\usepackage{amssymb}
\usepackage{amsthm}
\newtheorem{theorem}{Theorem}
\newtheorem{lemma}[theorem]{Lemma}
\newtheorem{definition}{Definition}
\newtheorem{corollary}[theorem]{Corollary}

\usepackage[hidelinks]{hyperref}
\newcommand{\R}{\mathbb{R}}

\usepackage{verbatim}
\usepackage{fancyvrb}
\usepackage{lineno}

\makeatletter
\def\ps@pprintTitle{%
	\let\@oddhead\@empty
	\let\@evenhead\@empty
	\def\@oddfoot{\centerline{\thepage}}%
	\let\@evenfoot\@oddfoot}
\makeatother

\begin{document}

\title{Identifying Super-Feminine, Super-Masculine and Sex-Defining Connections in the Human Braingraph}
	
\author[p]{László Keresztes\corref{cor2}}
\ead{keresztes@pitgroup.org}
\author[p]{Evelin Szögi\corref{cor2}}
\ead{szogi@pitgroup.org}
\author[p]{Bálint Varga}
\ead{balorkany@pitgroup.org}
\author[p,u]{Vince Grolmusz\corref{cor1}}
\ead{grolmusz@pitgroup.org}
\cortext[cor1]{Corresponding author}
\cortext[cor2]{Joint first authors}
\address[p]{PIT Bioinformatics Group, Eötvös University, H-1117 Budapest, Hungary}
\address[u]{Uratim Ltd., H-1118 Budapest, Hungary}

\date{}

\begin{abstract}
	For more than a decade now, we can discover and study thousands of cerebral connections with the application of diffusion magnetic resonance imaging (dMRI) techniques and the accompanying algorithmic workflow. While numerous connectomical results were published enlightening the relation between the braingraph and certain biological, medical, and psychological properties, it is still a great challenge to identify a small number of brain connections, closely related to those conditions. In the present contribution, by applying the 1200 Subjects Release of the Human Connectome Project (HCP), we identify just 102 connections out of the total number of 1950 connections in the 83-vertex graphs of 1065 subjects, which -- by a simple linear test -- precisely, without any error determine the sex of the subject. Very surprisingly, we were able to identify two graph edges out of these 102, if, whose weights, measured in fiber numbers, are all high, then the connectome always belongs to a female subject, independently of the other edges. Similarly, we have identified 3 edges from these 102, whose weights, if two of them are high and one is low, imply that the graph belongs to a male subject -- again, independently of the other edges. We call the former 2 edges superfeminine and the first two of the 3 edges supermasculine edges of the human connectome. Even more interestingly, one of the edges, connecting the right Pars Triangularis and the right Superior Parietal areas, is one of the 2 superfeminine edges, and it is also the third edge, accompanying the two supermasculine connections, if its weight is low; therefore it is also a ``switching'' connection.
\end{abstract}

\maketitle

\bigskip
\noindent Running head: Superfeminine \& Supermasculine Edges of the Connectome

\medskip
	
\section*{Introduction} 

One of the most most important challenges in brain science is establishing the cellular and anatomical causes of  neurophysiological or psychological differences between human subjects. In the last decade, by the spectacular developments in magnetic resonance imaging (MRI) of the brain, together with the data-processing pipeline for the data collected, our knowledge of the cerebral connections has been increased enormously (e.g., \cite{Sporns2005,VanEssen2012,Szalkai2016c}).

Diffusion MRI (dMRI) is capable of discovering the spatial anisotropy of the movement of water molecules in the brain: since in the axonal fibers of the white matter the water molecules have a diffusion movement along the axons, the axonal fibers can be tracked and traced, without any contrast material, with refined tractography algorithms \cite{Tournier2012}. With the reliable identification of the cortical- and sub-cortical gray matter areas \cite{Fischl2012}, we can construct the connectome, or the braingraph as follows: the nodes (or vertices) of this graph are the anatomically identified gray matter areas, and two nodes are connected by an (undirected) edge if the tractography algorithm finds axonal fibers between the brain areas, corresponding to these two nodes. 

Numerous results were published in the last decade, analyzing the human braingraph \cite{Hagmann2008,Szalkai2015a,Kerepesi2015,Hagmann2012,Szalkai2016d,Craddock2013a,Kerepesi2016,Szalkai2016,Ortiz2014}. Several works describe the connections of the healthy human brain \cite{Ball2014,Kerepesi2015b,Bargmann2012,Kerepesi2016b,Batalle2013,Szalkai2016e,Kerepesi2015c,Graham2014}, while others establish relations between psychiatric diseases or conditions and the connectome \cite{Agosta2014,AlexanderBloch2014,Baker2014,Szalkai2016c,Besson2014a,Bonilha2014}. 

\subsection*{Sex differences}

It is known for several years that the female and the male connectomes have different properties as graphs. The work of \cite{Ingalhalikar2014b} has proven -- on a publicly un-available dataset -- that the ratio of inter-hemispheric connections vs. the intra-hemispheric connections differs in males and females. 

Our group has shown on a publicly available dataset \cite{Kerepesi2016b} that several deep graph-theoretical properties, which are usually applied in the characterization of the quality of large computer interconnection networks \cite{Leighton2014}, are better in the braingraphs of women than in men \cite{Szalkai2015,Szalkai2016a}. We have proven that women's braingraphs are better expanders, have greater minimal bisection width, more spanning trees, larger minimum vertex cover than that of men. In the work of \cite{Szalkai2015c} we have proven that the advantage in the graph-quality parameters of women is due to the sex differences, and not to the size differences: we have compared the graphs of 36 large-brain women and 36 small-brain men, such that the brain volumes of all men were smaller than the brain volume of the smallest-brain woman in the group. We have found that men did not have better parameters than women in this test, and, additionally, many of the advantages of the women remained valid.

\subsection*{Parameters, defined {\em a priori} vs. {\em a posteriori}}

In the studies of \cite{Ingalhalikar2014b,Szalkai2015,Szalkai2016a,Szalkai2015c}, the authors compared parameters, which were identified {\em a priori}, i.e., the examination of these parameters were decided {\em before} the braingraphs were analyzed. In the present work, we intend to identify {\em a posteriori} parameters, i.e., edge-structures in the course of the analysis of the braingraphs, in which the male and female connectomes differ. Additionally, we intend to discover the smallest possible edge-sets of the braingraphs, which already determine the sex of the subject. 

First we constructed and trained a deep artificial neural network (ANN, see, e.g., \cite{Szalkai2017,Szalkai2017a} for definitions and examples) for classifying the sex of the subject, using only his/her braingraph. While these efforts were moderately successful, we have found that not the deep networks, but, on the contrary, the one level networks gave the best results for predicting the sex of the subject. In a certain sense, one-level neural networks are similar in their capabilities to simple linear test, or Support Vector Machines (SVMs). In the Methods section, we give a short introduction to SVMs. 

\subsection*{Few edges, which simply determine the sex of the subject}

Applying Support Vector Machines and integer programming algorithms, we were able to identify a small set of connectome edges, which precisely identify the female and male brains; and 2 and 3 particular edges, with the following property: if the fiber number of both edges are high enough, then the connectome belongs to a female subject. If the fiber number of the first two of the three edges are high, and the weight of the third is low enough, then the connectome belongs to a male subject. We call these edges superfeminine and supermasculine edges, respectively.

More exactly, we are considering graphs on 83 vertices. From these 83 vertices, one can form 
$${83\choose2}=3403$$
vertex-pairs, i.e., this is the maximum number of edges on 83 vertices. Note that each of the 1065 braingraphs contains exactly 83 vertices, and all of these vertices correspond to the very same 83 gray-matter areas of the brain (sometimes called ROIs, Regions of Interest). We consider the edges with weights, corresponding to the defining axonal fibers, individually scaled to a number between 0 and 1 (the details are given in the Methods section).

In our dataset of 1065 subjects, the  union of all the edges of the 1065 baingraphs contain 1950 edges. That means that out of the possible 3403 edges, only 1950 are present in the union of all the 1065 braingraphs. This is not a surprising observation since few areas from the left hemisphere are connected directly to the areas of the right hemisphere (see Supporting Figure 1 in the on-line supporting material).

If we consider these 1950 edges (or vertex-pairs, if the edge is not present) in any of our 1065 graphs, one can decide the exact identity of the particular braingraph (we note that no two subjects have exactly the same braingraph with same weights). Therefore, obviously, the set of these 1950 weighted edges defines {\em any} property of the subjects, including their sex. 

Consequently, it is not interesting that all edges define the sex (or other property) of the subject. However, it is a challenge to find the {\em smallest} possible set of the edges, which still implies the sex of the subject. This small set of connections may carry the most important features, which differentiate the braingraphs of the sexes.

We were able to identify 102 edges, which already determine the sex of the subjects (Fig. 1). Moreover, these edges determine the sex in a very simple, linear way, described below (the method of the identification of these 102 edges is detailed in the Methods section). For describing this phenomenon, let us correspond each graph to a length-102 vector, with coordinates equal to the edge-weights on the chosen 102 edges. This way, we have 1065 vectors, each with 102 coordinates. In other words, we have a 102-dimensional Euclidean space, with 1065 points (vectors) in it. In this space we have determined a hyperplane, which separates the male and female graphs in the following way: all the 102-dimensional vectors, made from the female graphs are on one side of the hyperplane, while all the 102-dimensional vectors, made from the male braingraphs are on the other side of the hyperplane. Consequently, (i)  102 edges out of the 1950 edges already determine the sex of the subject, and (ii) in a very simple, exact, and linear way, by a separating hyperplane.
Figure 1 gives a simple example for the data separation on the plane (in 2 dimensions) with a line (i.e., a line is a hyperplane on the plane).

\begin{figure}[H]
	\begin{center}
		\includegraphics[width=8cm]{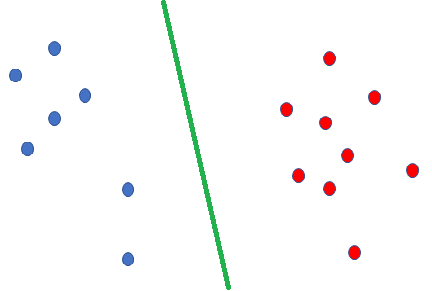}
		\caption{A simple example for Support Vector Machine data classification \cite{Cortes1995} on the plane. The blue and red points describe two classes of data (for example, each point corresponds to a braingraph, blue points to male, red points to female connectomes). The green line perfectly distinguishes the two classes: the blue ones are on one side, the red ones on the other side of the green line. In the 102-dimensional space (instead of the 2-dimensional space on the figure), we have succeeded to distinguish the male and female braingraphs in a similar way: all the male graphs are on one side, all the female graphs are on the other side of our hyperplane. The coordinates of the separating hyperplane are given in the Supporting material.}
	\end{center}
\end{figure}

 Figure 2 depicts the 102 edges, which already determine the sex of the subject. The list of these 102 edges is given in the Supporting Table 1. 

\begin{figure}[H]
	\begin{center}
		\includegraphics[width=12cm]{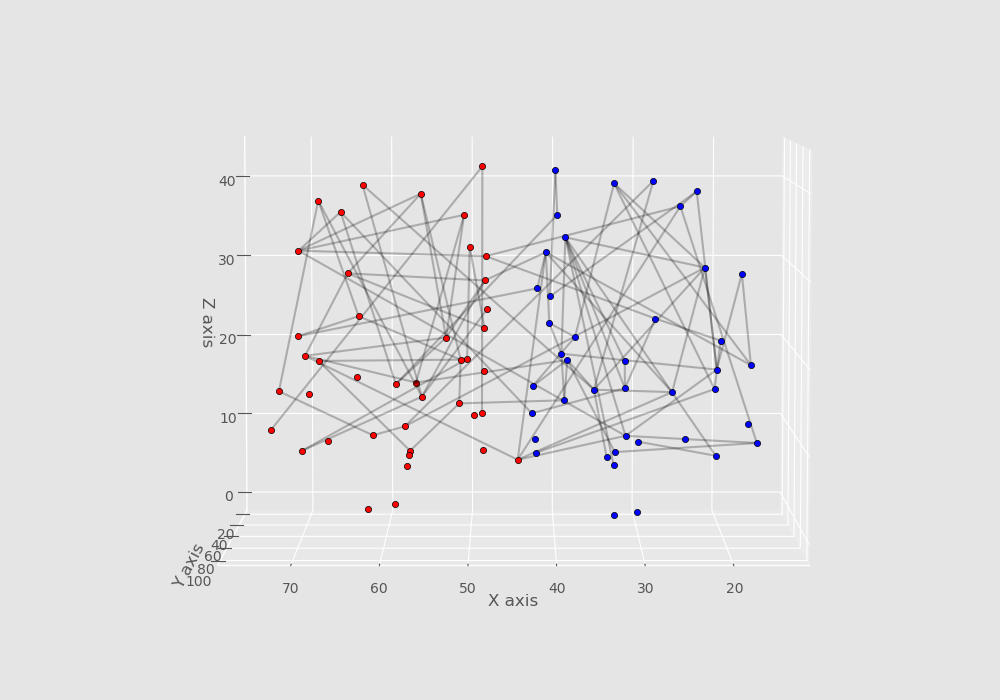}
		\caption{A braingraph of a subject, with 83 vertices and the 102 edges, whose weights (i.e., fiber numbers) already determine the sex of the subject. Labels on the axes are voxel coordinates in mm. In the 102-dimensional space, the male- and female braingraphs are perfectly separated by a hyperplane, similarly as the green line separates the blue and red dots on Figure 1.  The nodes from the distinct hemispheres are colored differently. The list of these 102 edges is given in Supporting Table 1 in the supporting material. }
	\end{center}
\end{figure}

\subsection*{Superfeminine and supermasculine edges}
	
Our second main result is the identification of very few connections, out of the 102 edges, in a way that if each of these edges has specific (either high or low) weights, then the sex of the subject is uniquely determined. 

Let us recall that the weight of an edge is the number of the axonal fibers found running between its two endpoints in the tractography algorithm, scaled for individual edges to be between 0 and 1 (the details are given in the Methods section). 

We have found that if the weights of both edges below are 1, then, independently from the weights of the remaining 100 edges out of the 102 sex-determining connections, the sex of the subject is female:

\begin{verbatim}
F1: (rh.superiorfrontal, Left-Putamen)
F2: (rh.parstriangularis, rh.superiorparietal)
\end{verbatim}

We call the set of edges F1. F2 ``superfeminine'' edges.

Similarly, we have found three edges, such that, if the weights of the first two are high and the weight of the third one is low, then, independently of the other edge-weights of the remaining 99 edges out of the 102 connections, the sex of the subject is male:

\begin{verbatim}
M1: (lh.rostralmiddlefrontal, Left-Thalamus-Proper)
M2: (Right-Hippocampus, lh.supramarginal)
F2: (rh.parstriangularis, rh.superiorparietal)
\end{verbatim}

The superfeminine and supermasculine edges are depicted on Figure 3.

We call edges M1 and M2  ``supermasculine'' edges. Note that edge F2 is present in both sets: if the weight of F1 and F2 are high, then it implies that the graph belongs to a female subject, and if the weight of F2 is low, and the weights of M1 and M2 are high, then the graph belongs to a male subject. We call the edge F2 a ``switching'' edge.

\section*{Methods}

\subsection*{Graph construction}

Our data source is the 1200 Subjects Release of the Human Connectome Project (HCP) \cite{McNab2013}, available at the \url{https://www.humanconnectome.org} site. The subjects were healthy adults between 22 and 35 years of age. We have applied the 3T MR diffusion imaging data and processed it with the Connectome Mapper Tool Kit (CMTK) \cite{Daducci2012}. 

Our goal was the construction of graphs, or connectomes, which describe the connections between the distinct, anatomically identified cortical and sub-cortical, gray-matter areas of the brain of the subjects. The nodes (or vertices) of our graphs corresponded to the anatomically identified gray matter areas, and we connected two nodes by an edge, if the workflow, described below, found axonal fibers, running between the areas that corresponded to the nodes. We emphasize that the study of the connectome instead of the whole MR image deals with {\em exclusively} the connections between the gray matter areas and does not take into account the exact orbit of the axonal fibers, running in the white matter of the brain. This way, we can work with graphs, instead of very redundant spatial imagery, gained from the processing of the diffusion MR images. We note that the (mathematical) graph theory, which was established in 1741 by a work of Euler \cite{Eulera}, has very rich structures and several of the most complex and deepest proofs and tools in mathematics (e.g., \cite{Szemeredi1975,Chudnovsky2006,Erdos1946}). Therefore, the transition from images to graphs facilitates the application of the well-developed techniques of the (mathematical) graph theory to one of the most complex organs on the Earth, the human brain.

The axonal fibers are discovered from the diffusion MR images by tractography algorithms. Probabilistic tractography was applied, with 1 million streamlines, by using MRtrix 0.2 tractography software. For each subject, the tractography program was run 10 times. In each run, the number of fibers was determined for each edge. If in any of the ten runs an edge was non-existent, that is, it was not defined by any fiber in the tractography, then that edge was discarded. Next, from these 10 runs, for each edge, the maximum and minimum number of fibers were deleted, and the average of the remaining 8 fiber numbers was assigned to the edge; this number is used as the weight of the edge. This way, the false positive and false negative edges were dealt with, and large errors, leading to the maximum or minimum fiber numbers of an edge, were discarded: they did not influence the average value.

 For each subject, 5 graphs, each with resolutions of 83, 129, 234, 463 and 1015 nodes were computed, by applying the CMTK's implementation of the FreeSurfer suite of programs for parcellation.  \cite{Fischl2012,Desikan2006,Tournier2012}.

The HCP public release contains the data of 1206 subjects. From these, 1113 contained structural scans. Our workflow was successfully completed for the data of 1065 subjects. The resulting graphs, with 5 resolutions for each subject, can be downloaded from the site \url{http://braingraph.org/download-pit-group-connectomes/}.

In the present work, we apply only the coarsest 83-node resolution, i.e., we consider 1065 graphs of 1065 subjects, each on 83 vertices. We have found 1950 edges by taking the union of the edges of the 1065 braingraphs on 83 vertices. 

In braingraph $u$ the edge $v$ is denoted by $e^u_v$, for $u=1,2,\ldots,1065$, $v=1,2,\ldots,1950$. The weight of the edge $e^u_v$, denoted by $w(e^u_v)$, is the average number of axonal fibers found running between its endpoints in the 8 tractography computations. 

\subsection*{An edge-specific weight-scaling method}

We would like to scale individually the weights of the edges such that all the resulting  edge-weights are between 0 and 1, as follows: 

$$x_{i}^\ell :=\frac{w(e_{i}^\ell) - \min\limits_{u=1}^k  w(e_{i}^u) }{\max\limits_{u=1}^k w(e_{i}^u) - \min\limits_{u=1}^k w(e_{i}^u)}\leqno{(1)}$$ if the denominator is not zero; otherwise, let $x_{i}^\ell$ be zero; $k=1065$. This way, for each braingraph, and for each edge, the smallest weight is transformed to 0, and the largest (if differs from the smallest) to 1. From now on, we use this scaled weights $x_i^\ell$, instead of the original ones. Let $x^\ell=(x_{1}^\ell,x_{2}^\ell,\ldots,x_{s}^\ell), s=1950$. 

In other words, for any $\ell$, $x^\ell$ describes a braingraph, with the new, scaled edges as its coordinates.

In what follows, we do not use the superscript $\ell$ if the meaning of $x$ is clear from the context.

\subsection*{An SVM-based technique with heuristic improvements}

The support vector machines (SVMs) are frequently used tools in artificial intelligence to classify the elements of large data sets \cite{Cortes1995}. 

Suppose that we have $k$ data points $x^1,x^2,\ldots, x^k$ in the $n$-dimensional Euclidean space $\R^n$, and a function $f:\R^n\to \{0,1\}$. We intend to find an $n$-dimensional hyperplane, such that (i) one side of the hyperplane contains all $x^i$'s with $f(x^i)=1$, and the other side of the hyperplane contains all $x^j$'s with $f(x^j)=0$ (ii) and the hyperplane separates the data points with the largest margin, that is, the distance of the closest data point to the hyperplane is maximized. 

If $n\geq k$ then the requirement (i) can always be met (one can see this simply by solving a linear systems of equations for finding the normal vector of the hyperplane). If $n<k$, then (i) (i.e., the perfect separation with a hyperspace) is not always satisfiable. We refer to Cover's theorem for probability estimations for the satisfiability of (i) when $n<k$ \cite{Cover1965}.

In the present work, first we solved (i) and (ii) for the $n=1950$ dimensional space, with $k=1065$, by using the Python Scikit-Learn suite. Next, we intend to reduce the coordinates (i.e., the number of edges), which are present in the separation. In other words, we needed to find as few coordinates as possible, such that the male and female connectomes can be separated by a hyperspace, using only the chosen coordinates.

This goal can be formalized as follows:

Let $\|w\|_0$ denote the number of the non-zero coordinates of vector $w$. Then we need to find

	$$ \min\|w\|_{0},\leqno{(2)}$$
 satisfying
	$$ w\cdot x + b \geq 0 \hbox{ for all } x,\leqno{(3)}$$ 
corresponding to a female braingraph, and
	$$ w\cdot x + b < 0  \hbox{ for all }x,\leqno{(4)}$$
corresponding to a male braingraph.
\medskip

By the best of our knowledge, no optimization method is known for solving this problem exactly in polynomial time. Here we have applied the combination of two simple heuristic solution methods, by which we were able to reduce $\|w\|_{0}$ from 1950 to 102. In other words, we can identify 102 coordinates of $x$ or, equivalently, 102 edges of the graph, such that the sex of the corresponding subject can be expressed by the sign of the linear expression $w\cdot x + b$ . The value of $b$ and the 102 non-zero coordinates of $w$ are given in the Supporting material, in Supporting Table 2.

The first heuristic algorithm is a Weight-Based Dimension-Reduction Algorithm (WBDRA):  Here, we start with a $w$, which separates linearly, and next delete of the smallest weight coordinates of $w$. A rate parameter $r$ defines that the $r$ fraction of the smallest coordinates needs to be deleted. If the new $w$ does not separate, then we backtrack and decrease $r$. The code of the algorithm is given in the Supporting Material, as Program Code 1.

The second procedure is a Single Dimension Deleting Algorithm (SDDA): Here we start with a separating $w$, and take a random order of the non-zero coordinates of $w$, and attempt to delete one dimension if the separation property remains valid. If not, then we try to delete the next dimension. The code of SDDA is given as Program Code 2 in the Supporting Material.

With the application of the three heuristic algorithms (WBDRA, SDDA, DDDA), we have succeeded in reducing the $\|w\|_{0}$ to 102.

We need to add that we cannot prove the optimality of the 102-dimensional solution: we think that even better results can be reached. However, by using Cover's theorem \cite{Cover1965}, the probability that randomly 0-1 labeled $k=1065$ points are separable by a hyperplane in 102 dimensions is much less than $2^{-100}$.

\subsection*{Finding Superfeminine and Supermasculine Edges}

Our goal is to identify edges, which have the greatest impact to the decisions (3) and (4). These edges may have very important roles in the sex-specific development and functioning of the human brain. Simply stated, the most important edges would have the coordinates with the largest absolute values in vector $w$ in (3) and (4). In what follows, we formally define 0-generator and 1-generator coordinates for a given function $f:[0,1]^N\to\{0,1\}$. 

Let $[N]$ denote the set $\{1, 2 ... N\}$. 

For  $y \in [0, 1]^{N}$ and $I \subset [N]$ let $y|_{I}\in[0, 1]^{N}$ denote:

\[y|_{I}(j)= \begin{cases} 
y_{j} & \textrm{ if $j \in I$} \\
0 & \textrm{ otherwise. } 
\end{cases}
\]

Let $\mathcal{G}$ denote the set of our 1065 braingraphs, each represented by an $x\in[0,1]^N$; originally, $N=1950$, i.e., each braingraph was represented by a 1950 weighted edges. In the previous section, we have seen that we can reduce $N=102$.

For an $I \subset [N]$ let $\mathcal{G}|_{I}:=\{x|_{I}: x \in \mathcal{G}\}$. 

\begin{definition}
	We say that $I \subset [N]$ is a $1$-generator for $f$ with a seed $x\in[0,1]^{N}|_{I}$, if $\forall y \in \mathcal{G}|_{[N]-I}$ $f(x+y) = 1$. Similarly, we say that $I \subset [N]$ is a $0$-generator for $f$ with a seed $x\in[0,1]^{N}|_{I}$ if $\forall y \in \mathcal{G}|_{[N]-I}$-re $f(x+y) = 0$
\end{definition}

In other words, the seed values in the coordinates in the 0-generator or 1-generator $I$ already determine the value of our $f$.

Our goal is finding the smallest 0- and 1-generators for $f$, where $f$ gives the sex of the subject: $f(x)=0$ for males, and $f(x)=1$ for females:

\[f(x)= \begin{cases} 
1 & \textrm{ if $w\cdot x + b \geq 0$} \\
0 & \textrm{ if $w\cdot x + b < 0$}
\end{cases}
\]

For this $f$, finding the minimal 0- and 1-generators is essentially a version of a knapsack problem, solvable by integer programming methods. For the reduction, we need some definitions and simple statements:

\begin{definition}
	Let $z_{F} \in [0, 1]^{N}$ be defined 
\[z_{F}(i)= \begin{cases} 
1 & \textrm{ if $w_{i} \geq 0$} \\
0 & \textrm{ if $w_{i} < 0$}
\end{cases}
\]
Let $z_{M} \in [0, 1]^{N}$ be defined 
\[z_{M}(i)= \begin{cases} 
1 & \textrm{ ha $w_{i} \leq 0$} \\
0 & \textrm{ ha $w_{i} > 0$}
\end{cases}
\]
\end{definition}

It is easy to see that $x=z_F$ maximizes and $x=z_M$ minimizes  $w\cdot x + b$.

We show the reduction for 1-generators, for 0-generators a similar reduction works.

\begin{lemma}
If $I \subset [N]$ is a $1$-generator for $f$ with seed $x\in[0,1]^{N}|_{I}$ then it is also a $1$-generator with seed $z_{F}|_{I}$ 
\end{lemma}

\begin{proof}
Let $y \in \mathcal{G}|_{[N]-I}$, then $w\cdot (z_{F}|_{I}+y) + b \geq w\cdot (x+y) > 0$.
\end{proof}

The next Corollary is obvious:

\begin{corollary}
	If $I$ the smallest 1-generator with any seed then it is also the smallest 1-generator with seed  $z_{F}|_{I}$.\qedsymbol
\end{corollary}

\begin{lemma}
 Let $\xi_{i}$ denote the coordinates of the 0-1 characteristic vector of set $I$: $\xi_{i}=1$ if and only if $i\in I$. Then $I$ is a 1-generator for $f$ with a seed $z_{F}|_{I}$ if and only if  $\forall x \in \mathcal{G}$, $f(x)=0$ implies $\sum_{i=1}^{N}\xi_{i}\cdot w_{i}(z_{F}(i)-x_{i}) > -w\cdot x - b$.
\end{lemma}

\begin{proof}
	$$\sum_{i=1}^{N}\xi_{i}\cdot w_{i}(z_{F}(i)-x_{i}) + w\cdot x + b = w\cdot (z_{F}|_{I}+x|_{[N]-I}) + b\leqno{(5)}$$ is non-negative. 
\end{proof}

Note that for any $x: f(x)=1$ (5) is also non-negative.

From Lemma 3, the optimization problem, which gives the minimum 1-generator, can be written:
Minimize $\sum_{i=1}^{N}\xi_{i}$, with the condition 
$$\sum_{i=1}^{N}\xi_{i}\cdot w_{i}(z_{F}(i)-x_{i}) > -w\cdot x - b.$$

We call the edges in 1-generators, where the corresponding seed coordinates are ones, superfeminine edges. We call the edges in 0-generators, where the corresponding seed coordinates are ones, supermasculine edges.

The distinction of by the seed-coordinates are made since the weights correspond to fiber numbers, and the ``strong'' graph edges, defined by many fibers, are called superfeminine or supermasculine edges. The superfeminine and supermasculine edges are depicted on Figure 3.

\begin{figure}[H]
	\begin{center}
		\includegraphics[width=12cm]{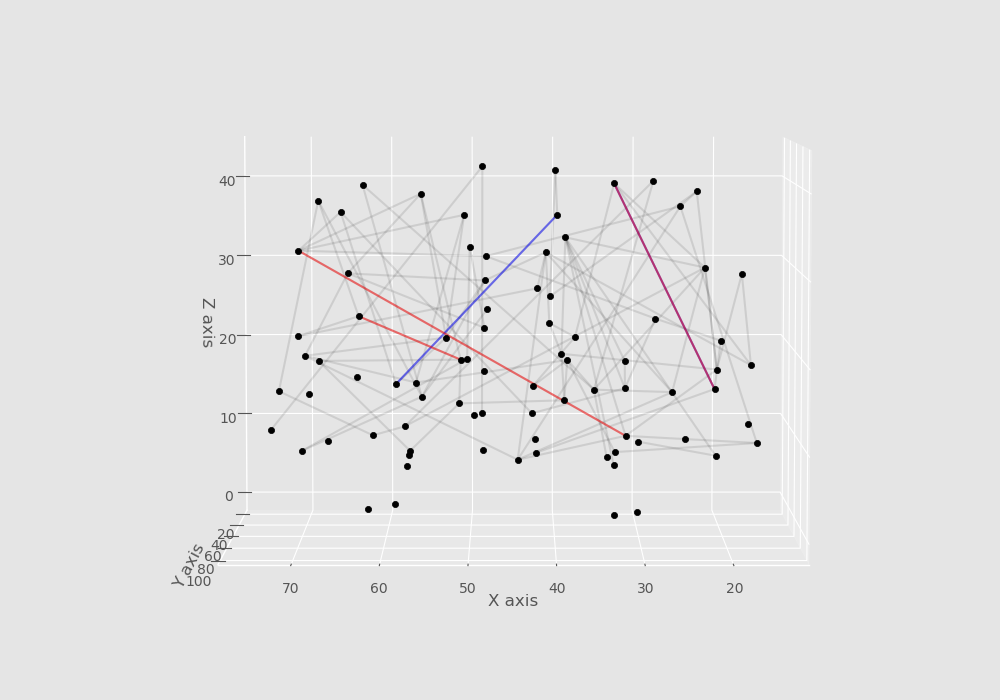}
		\caption{The superfeminine (blue) and supermasculine (red) edges. The switching edge is colored by purple. }
	\end{center}
\end{figure}

\subsection*{Software used} The braingraphs were computed by using the CMTK suite \cite{Daducci2012}, with the details given in the beginning of the section. The figures were created by using Python Matplotlib mplot3D and Networkx packages. The 1950-dimensional SVM was computed using the Python Scikit-Learn suite of programs. The heuristic improvements, resulting in the 102-dimensional separation, were found by the programs given in the Supporting Material in the Program codes section. For IP optimization we used the Python Pulp package.

\section*{Discussion and results}

Most cerebral sex dimorphisms studies to date were done on very small (up to 40-80 subjects) cohorts and applied mostly volumetric studies \cite{Frederikse1999,Koscik2009,Maleki2012,Butler2006}. Our previous works \cite{Szalkai2015,Szalkai2016a,Szalkai2015c,Fellner2017,Fellner2018,Fellner2019} first demonstrated sex dimorphisms in {\em a priori} defined graph parameters, in most cases the better connectivity-related parameters were found in the female connectomes.

Here we first demonstrate relatively small edge-sets, which determine the sex of the subjects on a very large, 1065-member cohort. 

The 102 edges, which already define the sex of the subjects are listed in the Supporting material as Supporting Table 1. Obviously, numerous edges connect subcortical nuclei with other parts of the brain. 13 of these 102 edges are inter-hemispheric. 

The most frequently appearing nodes in these 102 edges, without considering lateralization, are the inferiorparietal (10 times), posteriorcingulate (9 times), precuneus (9 times), superiorparietal (8 times). 

It is known that the inferior parietal lobule, which is a part of the heteromodal association cortex (HASC), shows sexual volumetric dimorphisms \cite{Frederikse1999,Koscik2009}. 

The sex differences in the development in migraine and the role of precuneus were reported in 
\cite{Maleki2012} and in mental rotation \cite{Butler2006}.

Counting with lateralization, the most frequent nodes are the rh.precuneus (7 times), rh.inferiorparietal (6 times), rh.posteriorcingulate (6 times) and the right-pallidum (6 times), all in the right hemisphere.

By the best of our knowledge, we are the firsts showing that not only these nodes of the braingraph, but rather their important connections, listed in the Supporting Table 1, carries substantial sex dimorphisms.   

Additionally, we are the firsts to show the existence of superfeminine and supermasculine edges. 

The superfeminine edges we have found are 

\begin{verbatim}
F1: (rh.superiorfrontal, Left-Putamen)
F2: (rh.parstriangularis, rh.superiorparietal).
\end{verbatim}

The two supermasculine edges with the F2 ``switching'' edge are:

\begin{verbatim}
M1: (lh.rostralmiddlefrontal, Left-Thalamus-Proper)
M2: (Right-Hippocampus, lh.supramarginal)
F2: (rh.parstriangularis, rh.superiorparietal)
\end{verbatim}

The most interesting edge is F2, which, with weight = 1, is a superfeminine edge, and with weight = 0, and with M1 and M2 with weights = 1, it implies the male sex of the subject. 

The area of Pars Triangularis was related to hormonal (oxytocin and arginine vasopressin) effects in men, and the same hormones to the parietal cortex -- instead of Pars Triangularis -- in women \cite{Rubin2017}. It is striking that just this edge, connecting the Pars Triangularis and the Superior Parietal area in the right hemisphere has this distinguished ``switching'' property.

There exists numerous other sets of edges with the superfeminne and supermasculine property, we demonstrated these since they were the smallest set we have found. We note that ours (F1, F2, and M1, M2) are relative to the 102-edge set, which we have identified as sex-defining ones. We note also that knowing only the weights of F1, F2 or M1, M2 and F1 will not imply the sex in general, only if they are 1,1 or 1,1,0 respectively.  

\section*{Conclusions}
 
First in the literature, we have followed an ``a posteriori'' way of search for edges in the human connectome, which determine the sex of the subjects. We have identified 102 edges that determine the sex in a very simple, linear way in a 1065-member cohort. Additionally, also first in the literature,  we have found two and three edges, out of the 102 ones, whose weights being properly set, imply the sex of the subject, independently of the other edges in the graph.

Our results were made possible by the novel edge-specific scaling of the weights of the edges, given by the formula (1). This scaling causes the otherwise low-weight edges also to strongly influence the linear tests (3) and (4). This effect is somewhat similar to the very successful relative PageRank protein-network ranking method in our previous work \cite{Banky2013}, where the low-degree network nodes have got chances to influence the node-ranking vector of the network.

\section*{Author contributions:}  LK and ES suggested using SVM in finding characteristic edges, performed integer linear programming optimizations and invented the methods of finding very few (i.e., 102) characteristic edges and the superfeminine and supermasculine edges. BV computed the braingraphs from the HCP public data. VG initiated the study, secured funding, analyzed the results, and wrote the paper. 

\section*{Data availability} The data source of this work was published at the Human Connectome Project's website at \url{http://www.humanconnectome.org} \cite{McNab2013} as the 1200-subjects public release. The parcellation data, containing the anatomically labeled ROIs, is listed in the CMTK nypipe GitHub repository \url{https://github.com/LTS5/cmp_nipype/blob/master/cmtklib/data/parcellation/lausanne2008/ParcellationLausanne2008.xls}. The braingraphs, computed by us, can be accessed at the  \url{http://braingraph.org/cms/download-pit-group-connectomes/} site. 

\section*{Acknowledgments}
Data were provided in part by the Human Connectome Project, WU-Minn Consortium (Principal Investigators: David Van Essen and Kamil Ugurbil; 1U54MH091657) funded by the 16 NIH Institutes and Centers that support the NIH Blueprint for Neuroscience Research; and by the McDonnell Center for Systems Neuroscience at Washington University. VG and BV were partially supported by the VEKOP-2.3.2-16-2017-00014 program, supported by the European Union and the State of Hungary, co-financed by the European Regional Development Fund, VG by the NKFI-126472 and NKFI-127909
grants of the National Research, Development and Innovation Office of Hungary. LK and ES were supported in part by the EFOP-3.6.3-VEKOP-16-2017-00002 grant, supported by the European Union, co-financed by the European Social Fund.
The authors are indebted to Balázs Szalkai for consultations on this work.
\bigskip 



\begin{thebibliography}{52}
	\providecommand{\natexlab}[1]{#1}
	\providecommand{\url}[1]{\texttt{#1}}
	\expandafter\ifx\csname urlstyle\endcsname\relax
	\providecommand{\doi}[1]{doi: #1}\else
	\providecommand{\doi}{doi: \begingroup \urlstyle{rm}\Url}\fi
	
	\bibitem[Sporns et~al.(2005)Sporns, Tononi, and K{\"{o}}tter]{Sporns2005}
	Olaf Sporns, Giulio Tononi, and Rolf K{\"{o}}tter.
	\newblock The human connectome: A structural description of the human brain.
	\newblock \emph{PLoS Computational Biology}, 1\penalty0 (4):\penalty0 e42, Sep
	2005.
	\newblock \doi{10.1371/journal.pcbi.0010042}.
	\newblock URL \url{http://dx.doi.org/10.1371/journal.pcbi.0010042}.
	
	\bibitem[{Van Essen} et~al.(2012){Van Essen}, Ugurbil, Auerbach, Barch,
	Behrens, Bucholz, Chang, Chen, Corbetta, Curtiss, {Della Penna}, Feinberg,
	Glasser, Harel, Heath, Larson-Prior, Marcus, Michalareas, Moeller,
	Oostenveld, Petersen, Prior, Schlaggar, Smith, Snyder, Xu, Yacoub, and
	~]{VanEssen2012}
	D.~C. {Van Essen}, K.~Ugurbil, E.~Auerbach, D.~Barch, T~E~J. Behrens,
	R.~Bucholz, A.~Chang, L.~Chen, M.~Corbetta, S.~W. Curtiss, S.~{Della Penna},
	D.~Feinberg, M.~F. Glasser, N.~Harel, A.~C. Heath, L.~Larson-Prior,
	D.~Marcus, G.~Michalareas, S.~Moeller, R.~Oostenveld, S.~E. Petersen,
	F.~Prior, B.~L. Schlaggar, S.~M. Smith, A.~Z. Snyder, J.~Xu, E.~Yacoub, and
	W.~U-Minn H. C. P~Consortium ~.
	\newblock The human connectome project: a data acquisition perspective.
	\newblock \emph{Neuroimage}, 62\penalty0 (4):\penalty0 2222--2231, Oct 2012.
	
	\bibitem[Szalkai et~al.(2019{\natexlab{a}})Szalkai, Varga, and
	Grolmusz]{Szalkai2016c}
	Balazs Szalkai, Balint Varga, and Vince Grolmusz.
	\newblock Mapping correlations of psychological and connectomical properties of
	the dataset of the human connectome project with the maximum spanning tree
	method.
	\newblock \emph{Brain Imaging and Behavior}, 13\penalty0 (5):\penalty0
	1185--1192, feb 2019{\natexlab{a}}.
	\newblock \doi{https://doi.org/10.1007/s11682-018-9937-6}.
	
	\bibitem[Tournier et~al.(2012)Tournier, Calamante, Connelly,
	et~al.]{Tournier2012}
	J~Tournier, Fernando Calamante, Alan Connelly, et~al.
	\newblock Mrtrix: diffusion tractography in crossing fiber regions.
	\newblock \emph{International Journal of Imaging Systems and Technology},
	22\penalty0 (1):\penalty0 53--66, 2012.
	
	\bibitem[Fischl(2012)]{Fischl2012}
	Bruce Fischl.
	\newblock Freesurfer.
	\newblock \emph{Neuroimage}, 62\penalty0 (2):\penalty0 774--781, 2012.
	
	\bibitem[Hagmann et~al.(2008)Hagmann, Cammoun, Gigandet, Meuli, Honey, Wedeen,
	and Sporns]{Hagmann2008}
	Patric Hagmann, Leila Cammoun, Xavier Gigandet, Reto Meuli, Christopher~J.
	Honey, Van~J. Wedeen, and Olaf Sporns.
	\newblock Mapping the structural core of human cerebral cortex.
	\newblock \emph{PLoS Biol}, 6\penalty0 (7):\penalty0 e159, Jul 2008.
	\newblock \doi{10.1371/journal.pbio.0060159}.
	\newblock URL \url{http://dx.doi.org/10.1371/journal.pbio.0060159}.
	
	\bibitem[Szalkai et~al.(2015{\natexlab{a}})Szalkai, Kerepesi, Varga, and
	Grolmusz]{Szalkai2015a}
	Bal{\'a}zs Szalkai, Csaba Kerepesi, B{\'a}lint Varga, and Vince Grolmusz.
	\newblock The {B}udapest {R}eference {C}onnectome {S}erver v2. 0.
	\newblock \emph{Neuroscience Letters}, 595:\penalty0 60--62,
	2015{\natexlab{a}}.
	
	\bibitem[Kerepesi and Grolmusz(2017)]{Kerepesi2015}
	Csaba Kerepesi and Vince Grolmusz.
	\newblock The {G}iant {V}irus {F}inder discovers an abundance of giant viruses
	in the {A}ntarctic dry valleys.
	\newblock \emph{Archives of Virology}, 162\penalty0 (6):\penalty0 1671--1676,
	2017.
	
	\bibitem[Hagmann et~al.(2012)Hagmann, Grant, and Fair]{Hagmann2012}
	Patric Hagmann, Patricia~E. Grant, and Damien~A. Fair.
	\newblock {MR} connectomics: a conceptual framework for studying the developing
	brain.
	\newblock \emph{Front Syst Neurosci}, 6:\penalty0 43, 2012.
	\newblock \doi{10.3389/fnsys.2012.00043}.
	\newblock URL \url{http://dx.doi.org/10.3389/fnsys.2012.00043}.
	
	\bibitem[Szalkai et~al.(2019{\natexlab{b}})Szalkai, Kerepesi, Varga, and
	Grolmusz]{Szalkai2016d}
	Balazs Szalkai, Csaba Kerepesi, Balint Varga, and Vince Grolmusz.
	\newblock High-resolution directed human connectomes and the consensus
	connectome dynamics.
	\newblock \emph{PLoS ONE}, 14\penalty0 (4):\penalty0 e0215473, September
	2019{\natexlab{b}}.
	\newblock URL \url{https://doi.org/10.1371/journal.pone.0215473}.
	
	\bibitem[Craddock et~al.(2013)Craddock, Milham, and LaConte]{Craddock2013a}
	R~Cameron Craddock, Michael~P. Milham, and Stephen~M. LaConte.
	\newblock Predicting intrinsic brain activity.
	\newblock \emph{Neuroimage}, 82:\penalty0 127--136, Nov 2013.
	\newblock \doi{10.1016/j.neuroimage.2013.05.072}.
	\newblock URL \url{http://dx.doi.org/10.1016/j.neuroimage.2013.05.072}.
	
	\bibitem[Kerepesi et~al.(2018{\natexlab{a}})Kerepesi, Varga, Szalkai, and
	Grolmusz]{Kerepesi2016}
	Csaba Kerepesi, Balint Varga, Balazs Szalkai, and Vince Grolmusz.
	\newblock The dorsal striatum and the dynamics of the consensus connectomes in
	the frontal lobe of the human brain.
	\newblock \emph{Neuroscience Letters}, 673:\penalty0 51--55, March
	2018{\natexlab{a}}.
	\newblock \doi{10.1016/j.neulet.2018.02.052}.
	
	\bibitem[Szalkai et~al.(2017{\natexlab{a}})Szalkai, Kerepesi, Varga, and
	Grolmusz]{Szalkai2016}
	Balazs Szalkai, Csaba Kerepesi, Balint Varga, and Vince Grolmusz.
	\newblock Parameterizable consensus connectomes from the {H}uman {C}onnectome
	{P}roject: The {B}udapest {R}eference {C}onnectome {S}erver v3.0.
	\newblock \emph{Cognitive Neurodynamics}, 11\penalty0 (1):\penalty0 113--116,
	feb 2017{\natexlab{a}}.
	\newblock \doi{http://dx.doi.org/10.1007/s11571-016-9407-z}.
	
	\bibitem[Ortiz et~al.(2014)Ortiz, Gorriz, Ramirez, and
	Salas-Gonzalez]{Ortiz2014}
	A~Ortiz, JM~Gorriz, Javier Ramirez, and Diego Salas-Gonzalez.
	\newblock Improving {MR} brain image segmentation using self-organising maps
	and entropy-gradient clustering.
	\newblock \emph{Information Sciences}, 262:\penalty0 117--136, 2014.
	
	\bibitem[Ball et~al.(2014)Ball, Aljabar, Zebari, Tusor, Arichi, Merchant,
	Robinson, Ogundipe, Rueckert, Edwards, and Counsell]{Ball2014}
	Gareth Ball, Paul Aljabar, Sally Zebari, Nora Tusor, Tomoki Arichi, Nazakat
	Merchant, Emma~C. Robinson, Enitan Ogundipe, Daniel Rueckert, A~David
	Edwards, and Serena~J. Counsell.
	\newblock Rich-club organization of the newborn human brain.
	\newblock \emph{Proc Natl Acad Sci U S A}, 111\penalty0 (20):\penalty0
	7456--7461, May 2014.
	\newblock \doi{10.1073/pnas.1324118111}.
	\newblock URL \url{http://dx.doi.org/10.1073/pnas.1324118111}.
	
	\bibitem[Kerepesi et~al.(2016)Kerepesi, Szalkai, Varga, and
	Grolmusz]{Kerepesi2015b}
	Csaba Kerepesi, Balazs Szalkai, Balint Varga, and Vince Grolmusz.
	\newblock How to direct the edges of the connectomes: Dynamics of the consensus
	connectomes and the development of the connections in the human brain.
	\newblock \emph{PLOS One}, 11\penalty0 (6):\penalty0 e0158680, June 2016.
	\newblock URL \url{http://dx.doi.org/10.1371/journal.pone.0158680}.
	
	\bibitem[Bargmann(2012)]{Bargmann2012}
	Cornelia~I. Bargmann.
	\newblock Beyond the connectome: how neuromodulators shape neural circuits.
	\newblock \emph{Bioessays}, 34\penalty0 (6):\penalty0 458--465, Jun 2012.
	\newblock \doi{10.1002/bies.201100185}.
	\newblock URL \url{http://dx.doi.org/10.1002/bies.201100185}.
	
	\bibitem[Kerepesi et~al.(2017)Kerepesi, Szalkai, Varga, and
	Grolmusz]{Kerepesi2016b}
	Csaba Kerepesi, Balazs Szalkai, Balint Varga, and Vince Grolmusz.
	\newblock The braingraph. org database of high resolution structural
	connectomes and the brain graph tools.
	\newblock \emph{Cognitive Neurodynamics}, 11\penalty0 (5):\penalty0 483--486,
	2017.
	
	\bibitem[Batalle et~al.(2013)Batalle, Mu{\~{n}}oz-Moreno, Figueras, Bargallo,
	Eixarch, and Gratacos]{Batalle2013}
	Dafnis Batalle, Emma Mu{\~{n}}oz-Moreno, Francesc Figueras, Nuria Bargallo,
	Elisenda Eixarch, and Eduard Gratacos.
	\newblock Normalization of similarity-based individual brain networks from gray
	matter {MRI} and its association with neurodevelopment in infants with
	intrauterine growth restriction.
	\newblock \emph{Neuroimage}, 83:\penalty0 901--911, Dec 2013.
	\newblock \doi{10.1016/j.neuroimage.2013.07.045}.
	\newblock URL \url{http://dx.doi.org/10.1016/j.neuroimage.2013.07.045}.
	
	\bibitem[Szalkai et~al.(2017{\natexlab{b}})Szalkai, Varga, and
	Grolmusz]{Szalkai2016e}
	Bal{\'a}zs Szalkai, B{\'a}lint Varga, and Vince Grolmusz.
	\newblock The robustness and the doubly-preferential attachment simulation of
	the consensus connectome dynamics of the human brain.
	\newblock \emph{Scientific Reports}, 7\penalty0 (16118), 2017{\natexlab{b}}.
	\newblock \doi{10.1038/s41598-017-16326-0}.
	
	\bibitem[Kerepesi et~al.(2018{\natexlab{b}})Kerepesi, Szalkai, Varga, and
	Grolmusz]{Kerepesi2015c}
	Csaba Kerepesi, Bal{\'a}zs Szalkai, B{\'a}lint Varga, and Vince Grolmusz.
	\newblock Comparative connectomics: Mapping the inter-individual variability of
	connections within the regions of the human brain.
	\newblock \emph{Neuroscience Letters}, 662\penalty0 (1):\penalty0 17--21,
	2018{\natexlab{b}}.
	\newblock \doi{10.1016/j.neulet.2017.10.003}.
	
	\bibitem[Graham(2014)]{Graham2014}
	Daniel~J. Graham.
	\newblock Routing in the brain.
	\newblock \emph{Front Comput Neurosci}, 8:\penalty0 44, 2014.
	\newblock \doi{10.3389/fncom.2014.00044}.
	\newblock URL \url{http://dx.doi.org/10.3389/fncom.2014.00044}.
	
	\bibitem[Agosta et~al.(2014)Agosta, Galantucci, Valsasina, Canu, Meani,
	Marcone, Magnani, Falini, Comi, and Filippi]{Agosta2014}
	Federica Agosta, Sebastiano Galantucci, Paola Valsasina, Elisa Canu, Alessandro
	Meani, Alessandra Marcone, Giuseppe Magnani, Andrea Falini, Giancarlo Comi,
	and Massimo Filippi.
	\newblock Disrupted brain connectome in semantic variant of primary progressive
	aphasia.
	\newblock \emph{Neurobiol Aging}, May 2014.
	\newblock \doi{10.1016/j.neurobiolaging.2014.05.017}.
	\newblock URL \url{http://dx.doi.org/10.1016/j.neurobiolaging.2014.05.017}.
	
	\bibitem[Alexander-Bloch et~al.(2014)Alexander-Bloch, Reiss, Rapoport, McAdams,
	Giedd, Bullmore, and Gogtay]{AlexanderBloch2014}
	Aaron~F. Alexander-Bloch, Philip~T. Reiss, Judith Rapoport, Harry McAdams,
	Jay~N. Giedd, Ed~T. Bullmore, and Nitin Gogtay.
	\newblock Abnormal cortical growth in schizophrenia targets normative modules
	of synchronized development.
	\newblock \emph{Biol Psychiatry}, Feb 2014.
	\newblock \doi{10.1016/j.biopsych.2014.02.010}.
	\newblock URL \url{http://dx.doi.org/10.1016/j.biopsych.2014.02.010}.
	
	\bibitem[Baker et~al.(2014)Baker, Holmes, Masters, Yeo, Krienen, Buckner, and
	{\"{O}}ng{\"{u}}r]{Baker2014}
	Justin~T. Baker, Avram~J. Holmes, Grace~A. Masters, B~T~Thomas Yeo, Fenna
	Krienen, Randy~L. Buckner, and Dost {\"{O}}ng{\"{u}}r.
	\newblock Disruption of cortical association networks in schizophrenia and
	psychotic bipolar disorder.
	\newblock \emph{JAMA Psychiatry}, 71\penalty0 (2):\penalty0 109--118, Feb 2014.
	\newblock \doi{10.1001/jamapsychiatry.2013.3469}.
	\newblock URL \url{http://dx.doi.org/10.1001/jamapsychiatry.2013.3469}.
	
	\bibitem[Besson et~al.(2014)Besson, Dinkelacker, Valabregue, Thivard, Leclerc,
	Baulac, Sammler, Colliot, Leh{\'{e}}ricy, Samson, and Dupont]{Besson2014a}
	Pierre Besson, Vera Dinkelacker, Romain Valabregue, Lionel Thivard, Xavier
	Leclerc, Michel Baulac, Daniela Sammler, Olivier Colliot, St{\'{e}}phane
	Leh{\'{e}}ricy, S{\'{e}}verine Samson, and Sophie Dupont.
	\newblock Structural connectivity differences in left and right temporal lobe
	epilepsy.
	\newblock \emph{Neuroimage}, 100C:\penalty0 135--144, May 2014.
	\newblock \doi{10.1016/j.neuroimage.2014.04.071}.
	\newblock URL \url{http://dx.doi.org/10.1016/j.neuroimage.2014.04.071}.
	
	\bibitem[Bonilha et~al.(2014)Bonilha, Nesland, Rorden, Fillmore, Ratnayake, and
	Fridriksson]{Bonilha2014}
	Leonardo Bonilha, Travis Nesland, Chris Rorden, Paul Fillmore, Ruwan~P.
	Ratnayake, and Julius Fridriksson.
	\newblock Mapping remote subcortical ramifications of injury after ischemic
	strokes.
	\newblock \emph{Behav Neurol}, 2014:\penalty0 215380, 2014.
	\newblock \doi{10.1155/2014/215380}.
	\newblock URL \url{http://dx.doi.org/10.1155/2014/215380}.
	
	\bibitem[Ingalhalikar et~al.(2014)Ingalhalikar, Smith, Parker, Satterthwaite,
	Elliott, Ruparel, Hakonarson, Gur, Gur, and Verma]{Ingalhalikar2014b}
	Madhura Ingalhalikar, Alex Smith, Drew Parker, Theodore~D. Satterthwaite,
	Mark~A. Elliott, Kosha Ruparel, Hakon Hakonarson, Raquel~E. Gur, Ruben~C.
	Gur, and Ragini Verma.
	\newblock Sex differences in the structural connectome of the human brain.
	\newblock \emph{Proc Natl Acad Sci U S A}, 111\penalty0 (2):\penalty0 823--828,
	Jan 2014.
	\newblock \doi{10.1073/pnas.1316909110}.
	\newblock URL \url{http://dx.doi.org/10.1073/pnas.1316909110}.
	
	\bibitem[Leighton(1992)]{Leighton2014}
	F~Thomson Leighton.
	\newblock \emph{Introduction to parallel algorithms and architectures: Arrays,
		trees, hypercubes}.
	\newblock Elsevier, 1992.
	\newblock ISBN 9781483221151.
	
	\bibitem[Szalkai et~al.(2015{\natexlab{b}})Szalkai, Varga, and
	Grolmusz]{Szalkai2015}
	Bal{\'{a}}zs Szalkai, B{\'{a}}lint Varga, and Vince Grolmusz.
	\newblock Graph theoretical analysis reveals: Women's brains are better
	connected than men's.
	\newblock \emph{PLoS One}, 10\penalty0 (7):\penalty0 e0130045,
	2015{\natexlab{b}}.
	\newblock \doi{10.1371/journal.pone.0130045}.
	\newblock URL \url{http://dx.doi.org/10.1371/journal.pone.0130045}.
	
	\bibitem[Szalkai et~al.(2016)Szalkai, Varga, and Grolmusz]{Szalkai2016a}
	Bal{\'a}zs Szalkai, B{\'a}lint Varga, and Vince Grolmusz.
	\newblock The graph of our mind.
	\newblock \emph{arXiv preprint arXiv:1603.00904}, 2016.
	
	\bibitem[Szalkai et~al.(2018)Szalkai, Varga, and Grolmusz]{Szalkai2015c}
	Bal{\'a}zs Szalkai, B{\'a}lint Varga, and Vince Grolmusz.
	\newblock Brain size bias-compensated graph-theoretical parameters are also
	better in women's connectomes.
	\newblock \emph{Brain Imaging and Behavior}, 12\penalty0 (3):\penalty0
	663--673, 2018.
	\newblock \doi{10.1007/s11682-017-9720-0}.
	\newblock URL \url{http://dx.doi.org/10.1007/s11682-017-9720-0}.
	
	\bibitem[Szalkai and Grolmusz(2017)]{Szalkai2017}
	Balazs Szalkai and Vince Grolmusz.
	\newblock Near perfect protein multi-label classification with deep neural
	networks.
	\newblock \emph{Methods (San Diego, Calif.)}, Jul 2017.
	\newblock ISSN 1095-9130.
	\newblock \doi{10.1016/j.ymeth.2017.06.034}.
	
	\bibitem[Szalkai and Grolmusz(2018)]{Szalkai2017a}
	Balazs Szalkai and Vince Grolmusz.
	\newblock {SECLAF}: A webserver and deep neural network design tool for
	hierarchical biological sequence classification.
	\newblock \emph{Bioinformatics}, 2018.
	\newblock URL \url{https://doi.org/10.1093/bioinformatics/bty116}.
	
	\bibitem[Cortes and Vapnik(1995)]{Cortes1995}
	Corinna Cortes and Vladimir Vapnik.
	\newblock Support-vector networks.
	\newblock \emph{Machine Learning}, 20\penalty0 (3):\penalty0 273--297, 1995.
	
	\bibitem[McNab et~al.(2013)McNab, Edlow, Witzel, Huang, Bhat, Heberlein,
	Feiweier, Liu, Keil, Cohen-Adad, Tisdall, Folkerth, Kinney, and
	Wald]{McNab2013}
	Jennifer~A. McNab, Brian~L. Edlow, Thomas Witzel, Susie~Y. Huang, Himanshu
	Bhat, Keith Heberlein, Thorsten Feiweier, Kecheng Liu, Boris Keil, Julien
	Cohen-Adad, M~Dylan Tisdall, Rebecca~D. Folkerth, Hannah~C. Kinney, and
	Lawrence~L. Wald.
	\newblock The {H}uman {C}onnectome {P}roject and beyond: initial applications
	of 300 m{T}/m gradients.
	\newblock \emph{Neuroimage}, 80:\penalty0 234--245, Oct 2013.
	\newblock \doi{10.1016/j.neuroimage.2013.05.074}.
	\newblock URL \url{http://dx.doi.org/10.1016/j.neuroimage.2013.05.074}.
	
	\bibitem[Daducci et~al.(2012)Daducci, Gerhard, Griffa, Lemkaddem, Cammoun,
	Gigandet, Meuli, Hagmann, and Thiran]{Daducci2012}
	Alessandro Daducci, Stephan Gerhard, Alessandra Griffa, Alia Lemkaddem, Leila
	Cammoun, Xavier Gigandet, Reto Meuli, Patric Hagmann, and Jean-Philippe
	Thiran.
	\newblock The connectome mapper: an open-source processing pipeline to map
	connectomes with {MRI}.
	\newblock \emph{PLoS One}, 7\penalty0 (12):\penalty0 e48121, 2012.
	\newblock \doi{10.1371/journal.pone.0048121}.
	\newblock URL \url{http://dx.doi.org/10.1371/journal.pone.0048121}.
	
	\bibitem[Euler(1741)]{Eulera}
	Leonhard Euler.
	\newblock Solutio problematis ad geometriam situs pertinentis.
	\newblock \emph{Commentarii Academiae Scientarum Imperialis Petropolitanae},
	8\penalty0 (1):\penalty0 128--140, 1741.
	\newblock URL \url{http://eulerarchive.maa.org//docs/originals/E053.pdf}.
	
	\bibitem[Szemeredi(1975)]{Szemeredi1975}
	Endre Szemeredi.
	\newblock {Regular Partitions of Graphs}.
	\newblock In \emph{Colloq. Internat. CNRS, Univ. Orsay, Orsay, 1976}, volume
	260. CNRS, 1975.
	
	\bibitem[Chudnovsky et~al.(2006)Chudnovsky, Robertson, Seymour, and
	Thomas]{Chudnovsky2006}
	Maria Chudnovsky, Neil Robertson, Paul Seymour, and Robin Thomas.
	\newblock The strong perfect graph theorem.
	\newblock \emph{Annals of Mathematics}, 164\penalty0 (1):\penalty0 51--229,
	2006.
	
	\bibitem[Erdos et~al.(1946)Erdos, Stone, et~al.]{Erdos1946}
	Paul Erdos, Arthur~H Stone, et~al.
	\newblock On the structure of linear graphs.
	\newblock \emph{Bull. Amer. Math. Soc}, 52\penalty0 (1087-1091):\penalty0 1,
	1946.
	
	\bibitem[Desikan et~al.(2006)Desikan, S{\'{e}}gonne, Fischl, Quinn, Dickerson,
	Blacker, Buckner, Dale, Maguire, Hyman, Albert, and Killiany]{Desikan2006}
	Rahul~S. Desikan, Florent S{\'{e}}gonne, Bruce Fischl, Brian~T. Quinn,
	Bradford~C. Dickerson, Deborah Blacker, Randy~L. Buckner, Anders~M. Dale,
	R~Paul Maguire, Bradley~T. Hyman, Marilyn~S. Albert, and Ronald~J. Killiany.
	\newblock An automated labeling system for subdividing the human cerebral
	cortex on mri scans into gyral based regions of interest.
	\newblock \emph{Neuroimage}, 31\penalty0 (3):\penalty0 968--980, Jul 2006.
	\newblock \doi{10.1016/j.neuroimage.2006.01.021}.
	\newblock URL \url{http://dx.doi.org/10.1016/j.neuroimage.2006.01.021}.
	
	\bibitem[Cover(1965)]{Cover1965}
	Thomas~M Cover.
	\newblock Geometrical and statistical properties of systems of linear
	inequalities with applications in pattern recognition.
	\newblock \emph{IEEE Transactions on Electronic Computers}, EC-14\penalty0
	(3):\penalty0 326--334, 1965.
	
	\bibitem[Frederikse et~al.(1999)Frederikse, Lu, Aylward, Barta, and
	Pearlson]{Frederikse1999}
	Melissa~E Frederikse, Angela Lu, Elizabeth Aylward, Patrick Barta, and Godfrey
	Pearlson.
	\newblock Sex differences in the inferior parietal lobule.
	\newblock \emph{Cerebral Cortex}, 9\penalty0 (8):\penalty0 896--901, 1999.
	
	\bibitem[Koscik et~al.(2009)Koscik, O'Leary, Moser, Andreasen, and
	Nopoulos]{Koscik2009}
	Tim Koscik, Dan O'Leary, David~J Moser, Nancy~C Andreasen, and Peg Nopoulos.
	\newblock Sex differences in parietal lobe morphology: relationship to mental
	rotation performance.
	\newblock \emph{Brain and cognition}, 69\penalty0 (3):\penalty0 451--459, 2009.
	
	\bibitem[Maleki et~al.(2012)Maleki, Linnman, Brawn, Burstein, Becerra, and
	Borsook]{Maleki2012}
	Nasim Maleki, Clas Linnman, Jennifer Brawn, Rami Burstein, Lino Becerra, and
	David Borsook.
	\newblock Her versus his migraine: multiple sex differences in brain function
	and structure.
	\newblock \emph{Brain : a journal of neurology}, 135:\penalty0 2546--2559,
	August 2012.
	\newblock ISSN 1460-2156.
	\newblock \doi{10.1093/brain/aws175}.
	
	\bibitem[Butler et~al.(2006)Butler, Imperato-McGinley, Pan, Voyer, Cordero,
	Zhu, Stern, and Silbersweig]{Butler2006}
	Tracy Butler, Julianne Imperato-McGinley, Hong Pan, Daniel Voyer, Juan Cordero,
	Yuan-Shan Zhu, Emily Stern, and David Silbersweig.
	\newblock Sex differences in mental rotation: top-down versus bottom-up
	processing.
	\newblock \emph{NeuroImage}, 32:\penalty0 445--456, August 2006.
	\newblock ISSN 1053-8119.
	\newblock \doi{10.1016/j.neuroimage.2006.03.030}.
	
	\bibitem[Fellner et~al.(2019{\natexlab{a}})Fellner, Varga, and
	Grolmusz]{Fellner2017}
	Mate Fellner, Balint Varga, and Vince Grolmusz.
	\newblock The frequent subgraphs of the connectome of the human brain.
	\newblock \emph{Cognitive Neurodynamics}, 13\penalty0 (5):\penalty0 453--460,
	2019{\natexlab{a}}.
	\newblock URL \url{https://doi.org/10.1007 /s11571-019-09535-y}.
	
	\bibitem[Fellner et~al.(2018)Fellner, Varga, and Grolmusz]{Fellner2018}
	Mate Fellner, Balint Varga, and Vince Grolmusz.
	\newblock The frequent network neighborhood mapping of the human hippocampus
	shows much more frequent neighbor sets in males than in females.
	\newblock \emph{arXiv preprint arXiv:1811.07423}, 2018.
	
	\bibitem[Fellner et~al.(2019{\natexlab{b}})Fellner, Varga, and
	Grolmusz]{Fellner2019}
	M{\'a}t{\'e} Fellner, B{\'a}lint Varga, and Vince Grolmusz.
	\newblock The frequent complete subgraphs in the human connectome.
	\newblock In \emph{International Work-Conference on Artificial Neural
		Networks}, volume 11507, pages 908--920. Springer, Springer,
	2019{\natexlab{b}}.
	
	\bibitem[Rubin et~al.(2017)Rubin, Yao, Keedy, Reilly, Bishop, Carter,
	Pournajafi-Nazarloo, Drogos, Tamminga, Pearlson, Keshavan, Clementz, Hill,
	Liao, Ji, Lui, and Sweeney]{Rubin2017}
	Leah~H Rubin, Li~Yao, Sarah~K Keedy, James~L Reilly, Jeffrey~R Bishop, C~Sue
	Carter, Hossein Pournajafi-Nazarloo, Lauren~L Drogos, Carol~A Tamminga,
	Godfrey~D Pearlson, Matcheri~S Keshavan, Brett~A Clementz, Scot~K Hill, Wei
	Liao, Gong-Jun Ji, Su~Lui, and John~A Sweeney.
	\newblock Sex differences in associations of arginine vasopressin and oxytocin
	with resting-state functional brain connectivity.
	\newblock \emph{Journal of neuroscience research}, 95:\penalty0 576--586,
	January 2017.
	\newblock ISSN 1097-4547.
	\newblock \doi{10.1002/jnr.23820}.
	
	\bibitem[B{\'a}nky et~al.(2013)B{\'a}nky, Iv{\'a}n, and Grolmusz]{Banky2013}
	D{\'a}niel B{\'a}nky, G{\'a}bor Iv{\'a}n, and Vince Grolmusz.
	\newblock Equal opportunity for low-degree network nodes: a pagerank-based
	method for protein target identification in metabolic graphs.
	\newblock \emph{PLoS One}, 8\penalty0 (1):\penalty0 e54204, 2013.
	
\end{thebibliography}

\section*{Supporting Material}

\section*{Supporting Figure} 

\setcounter{figure}{0}   
\renewcommand{\figurename}{Supporting Figure}
\begin{figure}[H]
	\begin{center}
		\includegraphics[width=12cm]{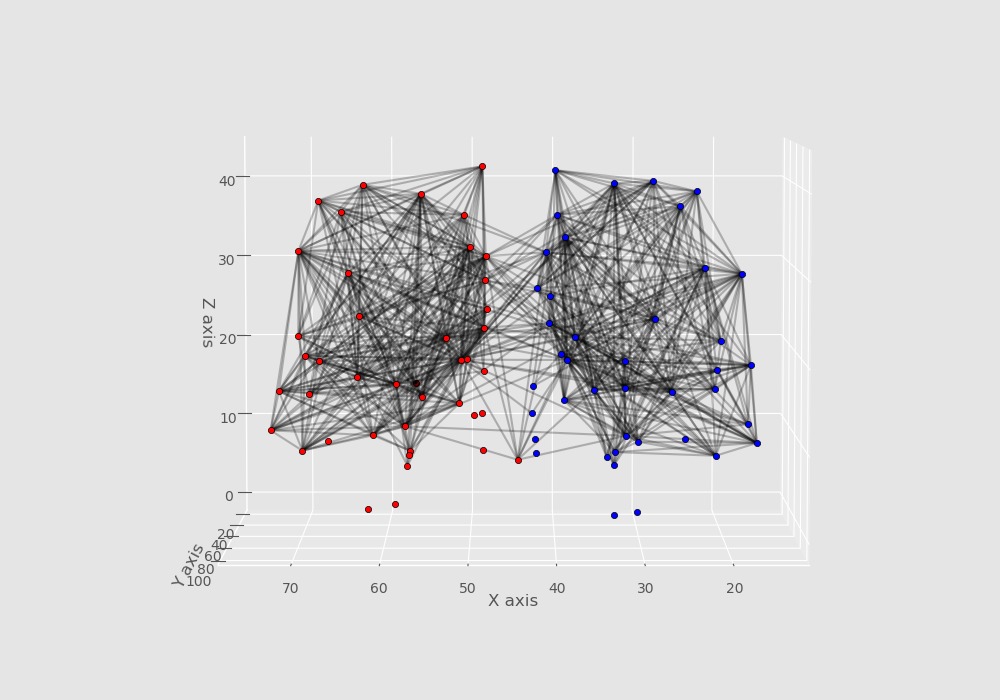}
		
		\caption{The braingraph of the subject of ID number 100206, with 83 vertices. The nodes from the left hemisphere are colored red, the ones from the right hemisphere are colored blue. }
	\end{center}
\end{figure}

\section*{Supporting Tables} 

\section*{Supporting Table 1}

Here we list the 102 edges we have identified characterizing the sex of the subjects. The ROIs are named according to the table given in  \url{https://github.com/LTS5/cmp_nipype/blob/master/cmtklib/data/parcellation/lausanne2008/ParcellationLausanne2008.xls}.

\begin{linenumbers}
	\begin{verbatim}
	(rh.precuneus, Right-Hippocampus)
	(rh.caudalmiddlefrontal, Left-Hippocampus)
	(rh.parahippocampal, Right-Thalamus-Proper)
	(rh.parsopercularis, rh.rostralmiddlefrontal)
	(rh.posteriorcingulate, rh.insula)
	(rh.posteriorcingulate, rh.bankssts)
	(lh.paracentral, Left-Accumbens-area)
	(rh.precuneus, rh.pericalcarine)
	(lh.rostralmiddlefrontal, Left-Thalamus-Proper)
	(rh.insula, Right-Pallidum)
	(rh.posteriorcingulate, Brain-Stem)
	(rh.inferiorparietal, rh.transversetemporal)
	(rh.lingual, lh.lingual)
	(lh.superiorparietal, Left-Caudate)
	(rh.precentral, lh.posteriorcingulate)
	(rh.parahippocampal, rh.middletemporal)
	(lh.rostralmiddlefrontal, lh.postcentral)
	(lh.precuneus, lh.pericalcarine)
	(rh.caudalanteriorcingulate, rh.posteriorcingulate)
	(lh.parsopercularis, lh.rostralmiddlefrontal)
	(lh.fusiform, lh.superiortemporal)
	(rh.inferiorparietal, rh.precuneus)
	(rh.superiorfrontal, Left-Putamen)
	(lh.caudalmiddlefrontal, Left-Pallidum)
	(lh.bankssts, Left-Caudate)
	(rh.paracentral, rh.cuneus)
	(rh.supramarginal, rh.bankssts)
	(rh.inferiorparietal, rh.insula)
	(rh.precentral, Brain-Stem)
	(Right-Hippocampus, Brain-Stem)
	(rh.lingual, Right-Thalamus-Proper)
	(lh.isthmuscingulate, lh.inferiorparietal)
	(rh.lateraloccipital, Right-Putamen)
	(lh.isthmuscingulate, lh.precuneus)
	(rh.pericalcarine, rh.lingual)
	(Right-Hippocampus, lh.supramarginal)
	(lh.postcentral, lh.superiortemporal)
	(rh.superiorparietal, rh.inferiorparietal)
	(lh.superiorparietal, Left-Thalamus-Proper)
	(lh.supramarginal, lh.superiorparietal)
	(rh.superiorparietal, Right-Caudate)
	(rh.middletemporal, Right-Hippocampus)
	(lh.pericalcarine, lh.inferiortemporal)
	(lh.posteriorcingulate, lh.supramarginal)
	(lh.transversetemporal, Left-Thalamus-Proper)
	(rh.precentral, rh.middletemporal)
	(rh.precentral, rh.postcentral)
	(rh.rostralmiddlefrontal, lh.posteriorcingulate)
	(rh.isthmuscingulate, rh.pericalcarine)
	(lh.superiorparietal, lh.inferiorparietal)
	(rh.caudalanteriorcingulate, lh.parsopercularis)
	(lh.inferiorparietal, lh.bankssts)
	(rh.parstriangularis, rh.parsopercularis)
	(rh.insula, Brain-Stem)
	(rh.precuneus, Right-Putamen)
	(lh.paracentral, lh.middletemporal)
	(rh.parstriangularis, rh.superiorparietal)
	(lh.inferiortemporal, Left-Pallidum)
	(rh.postcentral, rh.transversetemporal)
	(lh.caudalanteriorcingulate, Left-Pallidum)
	(rh.isthmuscingulate, Right-Caudate)
	(lh.fusiform, Left-Hippocampus)
	(Left-Caudate, Left-Putamen)
	(rh.lateralorbitofrontal, Right-Pallidum)
	(rh.superiorparietal, rh.bankssts)
	(lh.precentral, Left-Putamen)
	(lh.bankssts, Brain-Stem)
	(rh.precuneus, rh.lateraloccipital)
	(lh.caudalanteriorcingulate, lh.inferiorparietal)
	(Right-Putamen, Right-Accumbens-area)
	(lh.lingual, lh.parahippocampal)
	(Right-Pallidum, lh.caudalmiddlefrontal)
	(Right-Thalamus-Proper, Right-Pallidum)
	(rh.superiorfrontal, rh.paracentral)
	(rh.rostralanteriorcingulate, Right-Thalamus-Proper)
	(lh.lateraloccipital, lh.bankssts)
	(lh.caudalanteriorcingulate, Left-Caudate)
	(rh.supramarginal, rh.transversetemporal)
	(lh.superiorfrontal, lh.supramarginal)
	(lh.cuneus, Left-Pallidum)
	(rh.fusiform, rh.inferiortemporal)
	(rh.inferiorparietal, Right-Pallidum)
	(rh.rostralmiddlefrontal, Right-Putamen)
	(lh.superiorfrontal, lh.lateraloccipital)
	(rh.medialorbitofrontal, rh.parstriangularis)
	(lh.precentral, lh.supramarginal)
	(rh.transversetemporal, Right-Hippocampus)
	(Right-Thalamus-Proper, lh.lateraloccipital)
	(rh.posteriorcingulate, lh.caudalanteriorcingulate)
	(rh.inferiorparietal, Left-Hippocampus)
	(Right-Accumbens-area, lh.precentral)
	(rh.pericalcarine, rh.transversetemporal)
	(lh.parahippocampal, lh.transversetemporal)
	(rh.posteriorcingulate, rh.isthmuscingulate)
	(rh.rostralanteriorcingulate, Right-Caudate)
	(lh.lingual, Left-Thalamus-Proper)
	(rh.postcentral, rh.cuneus)
	(rh.caudalmiddlefrontal, Right-Pallidum)
	(lh.postcentral, Left-Pallidum)
	(lh.superiorfrontal, Left-Caudate)
	(rh.precuneus, Right-Amygdala)
	(rh.precuneus, rh.inferiortemporal)
	\end{verbatim}
\end{linenumbers}

\section*{Supporting Table 2}

Here we list the numerical values of the coefficients of the linear expression $w\cdot x + b$, which satisfies $$w\cdot x + b > 0$$ for all $x$, corresponding to a female braingraph, and
$$w\cdot x + b < 0$$ for all $x$, corresponding to a male braingraph.

The number $b=-6.038549870659588237e+01$. The coordinates of the 102-dimensional vector $w$, in the same order as the edges are listed in Supporting Table 1:

\resetlinenumber

\begin{linenumbers}
	\begin{verbatim}
	3.140827577736224896e+01
	-4.224516926777828019e+01
	-5.534731949947487095e+01
	-5.021383798706737167e+01
	3.338868535424175121e+01
	-6.441754383644450854e+01
	-9.137204014740284208e+00
	-1.623199041116811969e+01
	-1.404986301933692516e+02
	-4.006330872545306221e+01
	-4.252783335733649039e+01
	4.178817508618046617e+01
	4.372649839601213984e+01
	4.490601147649471869e+01
	4.707127337369107067e+01
	-3.636412443899504154e+01
	7.015011711412499551e+01
	4.336665095022959093e+01
	-7.814457117294969635e+01
	-4.214089662663334934e+01
	2.128444054185646195e+01
	3.312373784233776774e+01
	1.639952878714990732e+02
	3.026734348412119502e+01
	6.268856306230090070e+01
	-4.715689733082451340e+01
	2.432515011860931509e+01
	-3.448165961059984852e+01
	3.340173045666283969e+01
	4.042781804276076230e+01
	2.761020298703500586e+01
	2.402973906860708198e+01
	5.799632593673756986e+01
	2.811207336490517150e+01
	-3.530994163115634166e+01
	-1.569561312762774605e+02
	7.724860503993652117e+01
	-2.206056231378179433e+01
	2.931487757932982774e+01
	-2.102330615674154757e+01
	2.327980794756512850e+01
	7.887033703905483151e+01
	-1.652915811375330790e+01
	2.233205680678592842e+01
	-3.287713377723386543e+01
	3.098515708552118397e+01
	4.994991466114837664e+01
	-1.208699629202809689e+01
	-3.915248510214356514e+01
	-3.627962519972498256e+01
	-2.723952954348446909e+01
	-4.192812642391789524e+01
	-2.172699471903166213e+01
	-5.976128513966504840e+01
	6.259311751749017816e+01
	3.402550885586439477e+01
	2.210190354984765690e+02
	-3.676412998508114072e+01
	-8.721544084472516545e+01
	3.913733184781763441e+01
	3.834602562855815222e+01
	3.638168347536320368e+01
	4.718280018698974487e+01
	3.251867640180642383e+01
	4.918992981007945531e+01
	-2.066292878945655787e+01
	-4.066693930915500488e+01
	-2.157875932254564333e+01
	7.650679496763183352e+01
	4.825588522618672727e+01
	4.970415751004980365e+01
	7.295806396686961648e+01
	-2.882074964324494104e+01
	3.025214670489702584e+01
	7.392751412897777641e+01
	3.849139711312415812e+01
	-2.672208728717341941e+01
	-7.607620421231430896e+01
	-6.393366605226245980e+01
	2.138202760440539052e+01
	-3.614660559000751761e+01
	2.635867128635015533e+01
	7.575343108027709604e+01
	4.491884096447378738e+01
	9.919165525049973553e+01
	2.689606565844382047e+01
	-4.580873022162855079e+01
	3.428996344958203935e+01
	2.134704175565022766e+01
	-1.729583819036137982e+01
	5.079962213472451538e+01
	5.934224772754128452e+01
	5.915239459823803259e+01
	1.812222940561209938e+01
	5.407256405097979979e+01
	4.579401150729591308e+01
	7.263014051692987039e+01
	2.971896563339183928e+01
	6.289375521391245627e+01
	-4.140460314388891305e+01
	-4.235611871509448179e+01
	4.713752575052093619e+01\end{verbatim}
\end{linenumbers}

\section*{Program Codes}
\begin{sloppypar}

	\section*{Program Code 1}

\begin{Verbatim}
def drop_by_weight(features, rate=0.2, rate_divisor=1.2):
    tmp_features = features
    while True:
        x_ = x[:, tmp_features]
        clf = LinearSVC(random_state=0, tol=1e-5)
        clf.fit(x_, y)
        y_clf = clf.predict(x_)
        if np.array_equal(y, y_clf):
            features = tmp_features
            w_abs = np.sort(np.abs(clf.coef_[0]))
            threshold = w_abs[int(rate * len(w_abs))]
            tmp_features = [f for f in features if 
                  np.abs(clf.coef_[0, features.index(f)]) > threshold]
        else:
            rate_threshold = float(1)/len(features)
            if rate > rate_threshold:
                tmp_features = features
                rate = max(float(rate)/rate_divisor, rate_threshold)
            else:
                break

    return features
\end{Verbatim}
	
	\section*{Program Code 2}
	
\begin{Verbatim}
def drop_one(features):
    while True:
        fl = len(features)
        random.shuffle(features)
        for feature in features:
            clf = LinearSVC(random_state=0, tol=1e-5)
            tmp_features = list(features)
            tmp_features.remove(feature)
            clf.fit(x[:, tmp_features], y)
            y_clf = clf.predict(x[:, tmp_features])
            if np.array_equal(y, y_clf):
                features.remove(feature)
                break
        if len(features) == fl:
            break
    return features
\end{Verbatim}

\end{sloppypar}

\end{document}